\documentclass[a4paper]{amsart}
\usepackage{amsmath}
\usepackage{amssymb}
\usepackage{amsthm}
\usepackage{amsfonts}
\usepackage{amstext}
\usepackage{amsopn}
\usepackage{amsxtra}
\usepackage{graphicx}
\usepackage{pstricks}
\usepackage[latin1]{inputenc}
\usepackage{dsfont}
\usepackage{mathrsfs}
\usepackage{bm}

\theoremstyle{definition}

\theoremstyle{definition}

\theoremstyle{definition}

\newcommand{\gH}{\mathfrak{H}}

\newcommand{\cE}{\mathcal{E}}

\newcommand{\cP}{\mathcal{P}}
\newcommand{\cD}{\mathcal{D}}
\newcommand{\cT}{\mathcal{T}}
\newcommand{\cF}{\mathcal{F}}

\newcommand{\Z}{\mathbb{Z}}

\newcommand{\C}{\mathbb{C}}
\newcommand{\R}{\mathbb{R}}

\newcommand{\alp}{\boldsymbol{\alpha}}
\newcommand{\bA}{\boldsymbol{A}}
\newcommand{\bE}{\boldsymbol{E}}
\newcommand\ii{{\ensuremath {\infty}}}
\newcommand{\dK}{{\int_{-\infty}^{\infty}d\eta}}
\newcommand{\ide}{\frac 1{D^0 + i \eta}}

\newcommand{\hr}{{\widehat \rho}}
\newcommand{\alr}{{\alpha_{\rm ph}}}
\newcommand{\rr}{{\rho_{\rm ph}}}
\newcommand\pscal[1]{{\ensuremath{\left\langle #1 \right\rangle}}}

\newcommand{\CJ}{\mathscr{C}}
\renewcommand{\phi}{\varphi}
\def\tr{\mathop{\rm tr}\nolimits} 
\def\Tr{{\rm Tr}_{\C^4}}

\begin{document}

\title[A Minimization Method for Relativistic Electrons] {A Minimization Method for Relativistic Electrons in a Mean-Field Approximation of Quantum Electrodynamics}

\author[C. Hainzl]{Christian Hainzl}
\address{1300 University Boulevard, University of Alabama, Birmingham, USA} \email{hainzl@math.uab.edu}

\author[M. Lewin]{Mathieu Lewin}
\address{CNRS \& Université de Cergy-Pontoise, Laboratoire ``AGM'' UMR8088, 2 avenue Adolphe Chauvin, 95302 Cergy-Pontoise Cedex, FRANCE} \email{Mathieu.Lewin@math.cnrs.fr}

\author[E. S\'er\'e]{Eric S\'er\'e}
\address{Université Paris Dauphine, CEREMADE UMR7534, Place du Mar\'echal de Lattre de Tassigny, 75775 Paris
  Cedex 16, FRANCE.}
  \email{sere@ceremade.dauphine.fr}

\author[J-P Solovej]{Jan Philip Solovej}
\address{University of Copenhagen, Department of Mathematics,
 Universitetsparken 5, 2100 Copenhagen, DENMARK}
\email{solovej@math.ku.dk}

\begin{abstract}
We study a mean-field relativistic model which is able to describe both the behavior of finitely many spin-$1/2$
particles like electrons and of the Dirac sea which is self-consistently polarized in the presence of the real particles. The model is derived from the QED Hamiltonian in Coulomb gauge neglecting the photon field. 
All our results are non-perturbative and mathematically rigorous.
\end{abstract}

\maketitle

\tableofcontents


For heavy atoms, it is necessary to take relativistic effects into account. However there is no equivalent of the well-known $N$-body (non-relativistic) Schrödinger theory involving the Dirac operator, because of its negative spectrum.
The correct theory is Quantum Electrodynamics (QED). This theory has a remarkable predictive power but its
description in terms of perturbation theory restricts its range of applicability.
In fact a mathematically consistent formulation of the
nonperturbative theory is still unknown. On the other hand, effective models deduced from nonrelativistic theories (like the Dirac-Hartree-Fock model \cite{Sw}) suffer from inconsistencies: for instance a ground state never minimizes the physical energy which is always unbounded from below.

Here we study a \emph{variational} model based on a physical energy which can be minimized to obtain the ground state in a chosen charge sector. Our model describes the behavior of a finite number of particles (electrons), coupled to that of the Dirac sea which can become polarized. Although it plays a minor role in the calculation of the Lamb-shift for the ordinary hydrogen atom (comparing to other electrodynamic phenomena), vacuum polarization is important for High-$Z$ atoms \cite{MPS,Sh} and even plays a crucial role for muonic atoms \cite{FE,GRS}. We show that the introduction of the vacuum in the model is the solution to deal with the negative energies of the Dirac operator and obtain a well-defined ground state. This was predicted by Chaix and Iracane in \cite[page 3813]{CI}.

Our results are fully non-perturbative
and mathematically rigorous. The corresponding proofs are lengthy
and therefore published elsewhere \cite{HLS1,HLS2,HLSo,HLS3}.
Unfortunately, we have not yet been able to include the photon field in the model for mathematical reasons, but a model with photons can be formally written following our ideas. We emphasize that our goal is \emph{not} to obtain all the QED effects accurately but rather to show how the introduction of the self-consistent vacuum changes dramatically the general properties of the model, leading to a well-defined variational theory. The fact that optimal states are found by a minimization principle is important for computational purposes and is essential for a justification of Relativistic Density Functional Theory \cite{Engel,ED}.

Our methodology is as follows. We consider a Hartree-Fock type model in which particles interact through the Coulomb potential and with a kinetic energy given by the Dirac operator. Since we do not normal-order the underlying Hamiltonian, the kinetic energy is unbounded from below. However, we can as a first step construct the free Dirac sea by means of a thermodynamic limit. It is formally the minimizer of the Hartree-Fock energy. This state is not the usual sea of negative electrons of the free Dirac operator because all interactions between particles are taken into account, but it corresponds to filling negative energies of an effective mean-field translation-invariant operator. As a second step we introduce an external field potential and obtain a bounded-below energy by subtracting the (infinite) energy of the free self-consistent Dirac sea. In other words, we use the translation-invariant free vacuum as a reference and describe variations compared to it. We emphasize that this methodology is general and can be applied to other infinite quantum systems. It was used for the modelling of defects in crystals in \cite{CDL}.

\section{Formal derivation of the model}
We start with the \emph{formal} QED Hamiltonian written in Coulomb gauge, in the presence of an external electromagnetic potential $(V,a)$, see \cite{Hei,HE,Se,Sch1,BD}
\begin{multline}
\mathbb{H}^{V,a}= \int \Psi^*(x)\left[\alp\cdot(-i\nabla-\bA(x)-a(x))+m\beta\right]\Psi(x) \,dx+\int V(x)\rho(x)\,dx \\
+ \frac{\alpha}2 \iint \frac{\rho(x)\rho(y)}{|x-y|}dx\,dy+H_f
\label{Ham}
\end{multline}
In this formula, $\Psi(x)$ is the second quantized field operator which annihilates an electron at $x$ and satisfies the anticommutation relation
\begin{equation}
 \Psi^*(x)_\sigma\Psi(y)_\nu+\Psi(y)_\nu\Psi^*(x)_\sigma = 2\delta_{\sigma,\nu}\delta(x-y).
\label{CAR}
\end{equation}
The operator $\rho(x)$ is the \emph{density operator} defined by
\begin{equation}
\rho(x) =  \sum_{\sigma
=1}^4\frac{[\Psi_\sigma^*(x),\Psi_\sigma(x)]}{2}
\label{def_rho}
\end{equation}
where $[a,b]=ab-ba$. The operator $H_f$ describes the kinetic energy of the photons:
$$H_f=\frac{1}{8\pi\alpha}\int \left(|\nabla\times\bA(x)|^2+|\bE_t(x)|^2\right)\,dx=\frac{1}{\alpha}\sum_{\lambda=1,2}\int_{\R^3}dk\, |k|a^*_\lambda(k)a_\lambda(k)+\text{Cte}$$
(Cte indicates an infinite constant). 
The operators $\bA(x)$ and $\bE_t(x)$ are the electromagnetic field operators for the photons and $a^*_\lambda(k)$ is the creation operator of a photon with momentum $k$ and polarization $\lambda$.

In \eqref{Ham}, $(V,a)$ is an external electromagnetic potential, for instance created by a set of nuclei. We use the notation
$$D^0 = -i \alp \cdot \nabla + m \beta $$
for the Dirac operator. The constants $m$ and $\alpha$ appearing in \eqref{Ham} are respectively the (bare) mass and (bare) Sommerfeld fine structure constant for the electron. The units are chosen such that $\hbar=c=1$. The Hamiltonian $\mathbb{H}^{V,a}$ formally acts on the Fock space,
$$\cF=\cF_{\rm e}\otimes\cF_{\rm ph}$$
where $\cF_{\rm e}$ is the fermionic Fock space for the electrons and $\cF_{\rm ph}$ is the bosonic Fock space for the photons. 

We emphasize that \eqref{Ham} does not contain any normal-ordering or notion of (bare) electrons and positrons: $\Psi(x)$ can annihilate electrons of negative kinetic energy. The distinction between electrons and positrons should be a result of the theory and not an input. The commutator used in the formula \eqref{def_rho} of $\rho(x)$ is a kind of renormalization, independent of any reference. It is due to Heisenberg \cite{Hei} (see also \cite[Eq. $(96)$]{Pauli}) and it is necessary for a covariant formulation of QED, see \cite[Eq. $(1.14)$]{Sch1} and \cite[Eq. $(38)$]{Dy1}. More precisely, the Hamiltonian $\mathbb{H}^{V,a}$ possesses the interesting property of being invariant under charge conjugation since the following relations hold formally
$$\CJ\rho(x)\CJ^{-1}=-\rho(x),\qquad \CJ\mathbb{H}^{V,a} \CJ^{-1}=\mathbb{H}^{-V,a},$$
where $\CJ$ is the charge conjugation operator acting on the Fock space.

In our study of the QED Hamiltonian $\mathbb{H}^{V,a}$, we shall make two approximations:
\begin{itemize}
\item we neglect photons and assume there is no external magnetic field, $a\equiv0$;
\item we work in a mean-field theory, i.e. we restrict the Hamiltonian to Hartree-Fock states.
\end{itemize}
These approximations are of a different importance. Neglecting photons is of course a very rough approximation as it will forbid us to describe important physical effects occurring in QED like the self-energies of the electrons, the biggest contribution to the Lamb shift. But we do that only for mathematical reasons: we were not yet able to extend most of the results presented below when photons are taken into account. \emph{Formally}, a large part of our study is exactly the same with photons (when they are treated by a mean-field procedure). We hope to come back to this point in the near future.

The second approximation which we make by restricting ourselves to Hartree-Fock states is more fundamental and many of our results are specific to this case. Nevertheless, some of our general ideas may be applicable to the full QED model.

Let us recall that the electronic \emph{one-body density matrix} (two point function) of any electronic state $|\Omega\rangle\in\cF_{\rm e}$ is defined as
$$P(x,y)_{\sigma,\sigma'}=\langle\Omega|\Psi^*(x)_\sigma\Psi(y)_{\sigma'}|\Omega\rangle.$$
In view of \eqref{def_rho}, it is natural to introduce a \emph{renormalized one-body density matrix}
$$\gamma(x,y)_{\sigma,\sigma'}=\pscal{\Omega\left|\frac{[\Psi(x)_{\sigma}^*,\Psi(y)_{\sigma'}]}{2}\right|\Omega}.$$
By \eqref{CAR}, we obtain the simple relation
$$\gamma=P-\frac{I}{2}$$
where $I$ is the identity operator. Electronic Hartree-Fock states form a subset $\{|\Omega_P\rangle\}\subset\cF_{\rm e}$ of states which are completely determined by their density matrix $P$ (or equivalently by their renormalized density matrix $\gamma=P-I/2$). Recall that if 
$$|\Omega\rangle=|\phi_1\cdots\phi_N\rangle$$
is a Hartree-Fock states with $N$ occupied orbitals $\phi_1,...,\phi_N$, then the associated density matrix $P$ is just the orthogonal projector on ${\rm Span}(\phi_1,...,\phi_N)$:
$$P=\sum_{i=1}^N|\phi_i\rangle\langle\phi_i|.$$
For a \emph{formal} Hartree-Fock state with infinitely many occupied orbitals
$$|\Omega\rangle=|\phi_1\cdots\phi_N\cdots\rangle$$
we also obtain
$$P=\sum_{\rm occ}|\phi_i\rangle\langle\phi_i|.$$
Hence 
$$\gamma=P-\frac{I}2=\frac{P-P^\perp}{2}=\frac12\left(\sum_{\rm occ}|\phi_i\rangle\langle\phi_i|-\sum_{\rm unocc}|\phi_i\rangle\langle\phi_i|\right).$$
The associated density of charge is formally given by
\begin{equation}
\rho_\gamma(x)=\pscal{\Omega|\rho(x)|\Omega}=\frac12\left(\sum_{\rm occ}|\phi_i(x)|^2-\sum_{\rm unocc}|\phi_i(x)|^2\right). 
\label{rho_occ_unocc}
\end{equation}

Now we can compute the energy of any state $|\Omega_P\rangle\otimes|0\rangle$ where $|\Omega_P\rangle$ is a Hartree-Fock state in $\cF_{\rm e}$ and $|0\rangle\in\cF_{\rm ph}$ is the photonic vacuum. We obtain
$$\langle0|\otimes\langle\Omega_P|\mathbb{H}^{V,0}|\Omega_P\rangle\otimes|0\rangle=\cE_{\rm HF}^V(P-I/2)+\text{Cte}$$
where $\text{Cte}$ is an infinite constant and
\begin{multline}
 \cE_{\rm HF}^V(\gamma)=\tr(D^0\gamma)+\int V(x)\rho_\gamma(x)\,dx\\
+\frac{\alpha}{2}\iint\frac{\rho_\gamma(x)\rho_\gamma(y)}{|x-y|}dx\,dy
-\frac{\alpha}{2}\iint\frac{|\gamma(x,y)|^2}{|x-y|}dx\,dy.
\label{HF_QED}
\end{multline}
The reader can recognize the well-known Hartree-Fock energy, but applied to the renormalized density matrix $\gamma=P-I/2$ instead of the usual density matrix $P$. The last two terms of the first line are respectively the kinetic energy and the interaction energy of the electrons with the external potential $V$. In the second line appear respectively the so-called \emph{direct} and \emph{exchange} terms.
In Relativistic Density Functional Theory \cite{Engel, ED}, the exchange term is approximated by a function of $\rho_\gamma$ and its derivatives only.

Any stationary point of the above energy satisfies the first order equation (written in terms of the usual density matrix $P=\gamma+I/2$)
$$\left[P,F_{P-I/2}\right]=0$$
where $F_{P-I/2}$ is the Fock operator
$$F_{P-I/2}=D^0+V+\alpha\rho_{[P-I/2]}\ast\frac{1}{|x|}-\alpha\frac{(P-I/2)(x,y)}{|x-y|}.$$
For a minimizer (in a chosen charge sector), one will have the more precise equation
$$P=\chi_{(-\ii,\mu]}\left(F_{P-I/2}\right)$$
where $\mu$ is a Fermi level and $\chi_{(-\ii,\mu]}(A)$ is a mathematical notation for the spectral projector of $A$ corresponding to filling all energies $\leq\mu$. Saying differently, one obtains a Hartree-Fock state with infinitely many occupied orbitals, all having an energy $\leq\mu$. We shall give a precise interpretation of this equation later on.

\medskip

It is time to worry about the mathematical meaning of the formulas we have formally derived up to now, in particular the definition of the energy \eqref{HF_QED}. Unfortunately, the latter does not make any sense for the following reason: when $P$ is an orthogonal projector (as this is the case for HF states), $\gamma=P-I/2$ is never a compact operator in an infinite dimension space. Hence none of the terms appearing in \eqref{HF_QED} has a clear mathematical meaning. Formally, one has $\cE_{\rm HF}^V(P-I/2)=-\ii$ for any density matrix $P$.

In \cite{HLSo}, we proposed to overcome this difficulty in the following way: we restrict the whole system to a box of size $L$ with periodic boundary conditions and an ultraviolet cut-off $\Lambda$ in the Fourier domain. Then all the above formulas make perfectly sense because we are in a finite-dimensional setting. In particular one can define minimizers of the HF energy with or without the external field $V$, with or without a charge constraint. Then, we look at the limit of the minimizer in the considered class when the size of the box grows, $L\to\ii$, but the cut-off $\Lambda$ stays fixed. The limit (if it exists) is the formal minimizer of the unbounded below energy $\cE^V_{\rm HF}$.

Actually we shall essentially use this method to define the free vacuum (the global minimizer of $\cE^0_{\rm HF}$ when $V=0$). Once the free vacuum has been found, we formally subtract its (infinite) energy to the expression \eqref{HF_QED} and obtain a well-defined bounded below energy called \emph{Bogoliubov-Dirac-Fock} and which is related to a work of Chaix and Iracane \cite{CI}.

Notice the ultraviolet cut-off $\Lambda$ is fixed during the whole study. It is only at the very end that we can tackle the difficult task to remove it by \emph{renormalization}. We shall also discuss the appearence of the Landau pole.

We explain all that in details in the next sections.

\section{Restriction of the system to a box and definition of the free vacuum}
Let us consider a box of size $L$, $C_L:=[-L/2;L/2)^3$ and limit the system to this box, with periodic boundary conditions. For simplicity, we also periodize the Coulomb potential and introduce
\begin{equation}\label{Coutor}
G_L(x)=\frac 1{L^3}\left(\sum_{k \in \frac{2\pi \Z^3}{L}
\setminus \{0\}} \frac{4\pi}{|k|^2}e^{i kx} + cL^2\right)
\end{equation}
where $c$ is chosen such that $G\geq0$. Furthermore, we add an ultraviolet cut-off $\Lambda$, i.e. we choose as one-body space the finite-dimensional
$$\gH_\Lambda^L:={\rm Span}\left(e^{ik\cdot x}\ \bigg|\ k\in\frac{2\pi\Z^3}{L},\quad |k|\leq\Lambda\right).$$
The periodic Hartree-Fock energy (without external field $V$) is then defined as
\begin{multline}\label{freetorfunct}
\cE^0_L(\gamma_L)= \tr(D^0\gamma_L) + \frac \alpha 2 \iint
\rho_{\gamma_L}(x) G_L(x-y)\rho_{\gamma_L}(y) dxdy \\- \frac
\alpha 2 \iint {|\gamma_L(x,y)|^2}{G_L(x-y)}dx dy.
\end{multline}
This expression is well defined for all renormalized density matrices $\gamma_L$ acting on the one-body space $\gH_\Lambda^L$ and satisfying the condition that $\gamma_L+I/2$ is an orthogonal projector. Indeed, following a method of Lieb \cite{Lieb}, we can even relax this condition and work under the assumption that
\begin{equation}
 -I/2\leq\gamma_L\leq I/2.
\label{relaxation}
\end{equation}
It is possible to define the QED Hamiltonian without photons in the box in the same way, see \cite{HLSo}. Notice the fermionic Fock space built on the one-body space $\gH_\Lambda^L$ is also finite-dimensional.

The minimizing problem defining the free HF vacuum in the box reads
$$E_L^0:=\inf_{-I/2\leq\gamma_L\leq I/2}\cE^0_L(\gamma_L).$$
It was shown in \cite[Thm 2.7]{HLSo} that for $L\gg1$ and $0\leq\alpha<4/\pi$ this problem admits a \emph{unique minimizer} $\gamma^0_L$, which has several interesting properties. First it takes the form $\gamma^0_L=P^0_L-I/2$ where $P^0_L$ is an orthogonal projector acting on $\gH_\Lambda^L$, hence the relaxation \eqref{relaxation} does not change the minimum. Then, $\gamma^0_L$ is a translation-invariant operator, meaning that it is a multiplication operator in the Fourier domain, $\gamma^0_L=\gamma^0_L(k)$. It can also be proved that the associated density of charge vanishes, $\rho_{\gamma^0_L}\equiv0$. This comes from the fact that $\gamma^0_L$ has a very special form which we do not detail as we are more interested in the properties of the limit of $\gamma^0_L$ as $L\to\ii$.

Indeed, it was shown in \cite[Thm 2.7]{HLSo} that 
$$\gamma^0_L\to\gamma^0$$
uniformly as functions of the Fourier variable and that
\begin{equation}
 \frac{E^0_L}{L^3}\to \bar e
\label{thermo_limit_free_vac}
\end{equation}
as $L\to\ii$. The operator $\gamma^0$ is the density matrix of the free Hartree-Fock vacuum in the whole space (with the ultraviolet cut-off $\Lambda$), which formally minimizes the no-photon QED Hartree-Fock energy $\cE_{\rm HF}^0$ in
spite of the fact that its energy is $-\infty$.

We now describe the interesting properties of $\gamma^0$, which were proved in \cite[Thm 2.2]{HLSo} and \cite{LS}. First $\gamma^0=\gamma^0(p)$ is a translation-invariant operator acting on the one-body space
$$\gH_\Lambda:=\left\{f\in L^2(\R^3,\C^4),\ {\rm Supp}(\widehat{f})\subseteq B(0,\Lambda)\right\}$$
of functions whose Fourier transform is supported in the ball of radius $\Lambda$ (the natural ``limit'' of $\gH_\Lambda^L$). One has $\gamma^0=\cP^0_--I/2$ where $\cP^0_-$ is an orthogonal projector which satisfies the following SCF equation:
\begin{equation}
\left\{ \begin{array}{l}
\cP^0_-=\chi_{(-\ii;0]}(\cD^0),\\
\cD^0=D^0-\alpha\frac{(\cP^0_--I/2)(x-y)}{|x-y|}.
\end{array}\right.
\label{SCF_free_vac}
\end{equation}
The operator $\cD^0$ is the self-consistent (SCF) Fock operator of the free vacuum $\cP^0_-$. It was shown that it takes the following special form
$$\cD^0(p)=g_1(|p|)\alp\cdot p+g_0(|p|)\beta$$
with
$$1\leq g_1(|p|)\leq \frac{g_0(|p|)}{m},$$
hence
\begin{equation}
 |D^0(p)|\leq |\cD^0(p)|=\sqrt{g_1(|p|)^2|p|^2+g_0(|p|)^2},
\label{prop_D_0}
\end{equation}
i.e. the gap of $\cD^0$ is bigger than the one of the original Dirac operator $D^0$. Notice \eqref{SCF_free_vac} corresponds to the usual Dirac's picture that the free vacuum is a Hartree-Fock state occupying all the negative energies of a Dirac-type operator. If $\alpha=0$ (no interaction), then we get the original picture $\cP^0_-=P^0_-:=\chi_{(-\ii;0]}(D^0)$, but in general $\cP^0_-\neq P^0_-$.

\begin{figure}[h]
\input{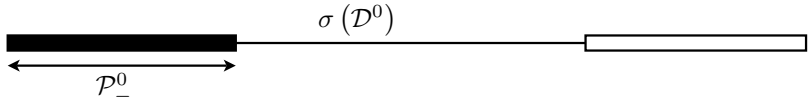}
\caption{The free vacuum $\cP^0_-$ fills the negative energies of the SCF Fock operator $\cD^0$.}
\end{figure}

Notice \eqref{SCF_free_vac} can be rewritten in terms of $\gamma^0$ in the form 
\begin{equation}
\gamma^0(p)=-\frac{\cD^0(p)}{2|\cD^0(p)|} = -\frac{g_1(|p|)}{2\sqrt{g_1(|p|)^2|p|^2+g_0(|p|)^2}}\alp\cdot p-\frac{g_0(|p|)}{2\sqrt{g_1(|p|)^2|p|^2+g_0(|p|)^2}}\beta.
\label{form_gamma_0} 
\end{equation}
In QED, the Feynman propagator at equal times
$$S_F(x,y;t_x=t_y):=i\gamma(x,y)\beta$$
is often expressed using the Källén-Lehmann representation \cite{Ka,Leh,BD}, based on relativistic invariances. Although our model is not fully relativistically invariant (we discard photons and use an ultraviolet cut-off $\Lambda$) and is only defined in the mean-field approximation, our solution \eqref{form_gamma_0} has exactly the form which may be derived from the Källén-Lehmann representation for the equal time propagator. In four-dimensional full QED, a self-consistent equation similar to \eqref{SCF_free_vac} is well-known and used. These so-called Schwinger-Dyson equations \cite{Sch4,Dy2} have been approximately solved for the free vacuum case first by Landau {\it et al.} in \cite{Lan81,Lan89}, and then by many authors (see, e.g., \cite{mass1,mass2,mass3}). Equation \eqref{SCF_free_vac} has already been studied by Lieb and Siedentop in \cite{LS} in a different setting.

We notice that
$$\rho_{\gamma^0}\equiv0.$$
This is indeed a consequence of Formula \eqref{form_gamma_0}: one has $C\gamma^0C^{-1}=-\gamma^0$ where $C$ is the charge conjugation operator. Hence any negative energy state of $\gamma^0$ can be associated to a positive energy state obtained by charge conjugation. The result follows from \eqref{rho_occ_unocc}. In mathematical terms,
$\rho_{\gamma^0}(x)=\tr_{\C^4}\gamma^0(x,x)=0,$
the Dirac matrices being trace-less.

We want to mention a last interesting property of $\gamma^0$: it is indeed the \emph{unique minimizer} of the energy per unit volume defined by
\begin{equation}
\cT(\gamma)=\frac{1}{(2\pi)^{3}}\int_{|p|\leq\Lambda}\tr_{\C^4}[D^0(p)\gamma(p)]dp-\frac{\alpha}{(2\pi)^5}\iint_{|p|,|q|\leq\Lambda}\frac{\tr_{\C^4}[\gamma(p)\gamma(q)]}{|p-q|^2}dp\,dq,
\label{def_fn_F0}
\end{equation} 
where we recall that $\Lambda$ is the ultraviolet cut-off. This property can be used for the numerical computation of the free vacuum $\gamma^0$. Lastly, we have that the energy per unit volume of the free vacuum is
$$\bar e=\inf_{\substack{\gamma=\gamma(p),\\ -I/2\leq\gamma\leq I/2}}\cT(\gamma),$$
the limit appearing in \eqref{thermo_limit_free_vac} as proved in \cite[Thm 2.7]{HLSo}.

\section{Bogoliubov-Dirac-Fock Theory}
If we summarize, using a thermodynamic limit we have been able to define the free vacuum which is the unique minimizer of $\cE_{\rm HF}^0$, and despite the fact that its energy is $-\ii$. The free vacuum is the negative Dirac sea of an SCF translation-invariant Dirac Fock operator $\cD^0$. Now we use this vacuum as a reference and subtract its (infinite) energy to the original HF energy, in order to obtain a bounded-below function. Formally, this gives for any state $\gamma=P-I/2$ the so-called \emph{Bogoliubov-Dirac-Fock (BDF) energy} \cite{CI,Chaix}
\begin{align}
 \cE^V_{\rm BDF}(P-\cP^0_-) & := \langle0|\otimes\langle\Omega_P|\mathbb{H}^{V,0}|\Omega_P\rangle\otimes|0\rangle-\pscal{\Omega_0|\mathbb{H}^{0,0}|\Omega_0}\nonumber\\
 & = \cE^V_{\rm HF}(P-I/2)-\cE^0_{\rm HF}(\cP^0_--I/2)\nonumber\\
 &= \tr\cD^0(P-\cP_-^0)+\int V(x)\rho_{P-\cP^0_-}(x)\,dx\nonumber\\
& \quad +\frac{\alpha}{2}\iint\frac{\rho_{P-\cP_-^0}(x)\rho_{P-\cP_-^0}(y)}{|x-y|}dx\,dy
-\frac{\alpha}{2}\iint\frac{|(P-\cP_-^0)(x,y)|^2}{|x-y|}dx\,dy\label{comput_BDF}
\end{align}
where $|\Omega_0\rangle=|\Omega_{\cP^0_-}\rangle\otimes|0\rangle$ is the no-photon HF free vacuum in Fock space found in the previous section. In \eqref{comput_BDF} we have used that $\rho_{\cP^0_--I/2}\equiv0$ and recognized the formula \eqref{SCF_free_vac} of $\cD^0$. 

The BDF energy measures the energy of any state $\gamma$ compared to the (infinite) energy of the free vacuum $\gamma^0$. Also $Q=P-\cP^0_-$ describes the variations counted with respect to the free Dirac sea. The BDF energy was first introduced by Chaix and Iracane \cite{CI} but with $\cP^0_-$ and $\cD^0$ replaced by $P^0_-$ and $D^0$. It was first mathematically studied in \cite{BBHS}. Chaix and Iracane obtained their energy by imposing from the beginning a normal-ordering on the QED Hamiltonian, taking as definition of positrons and electrons the ones given by the decomposition induced by $P^0_-$. If $\alpha=0$ our model is equivalent to the one of Chaix-Iracane, but it is not when $\alpha\neq0$. It seems that normal ordering is only fully relevant for the description of non interacting systems.

Once again the above formal computation \eqref{comput_BDF} can be justified by a thermodynamic limit. We will show that the last expression of $\cE_{\rm BDF}^V$ is well-defined mathematically and we will be able to find minimizers of this energy. We can prove that any sequence of minimizers in boxes will converge to these states in the thermodynamic limit $L\to\ii$ but we do not give more details and refer to \cite[Thm 2.9]{HLSo}.

\medskip

We now explain how it is possible to give a mathematical meaning to the last expression of \eqref{comput_BDF}. Some details which may appear as mathematical technicalities will later reveal to be crucial for renormalization, hence related to important physical properties. We recall that an operator $Q$ is said to be \emph{trace-class} when $\sum_i\pscal{\phi_i|\sqrt{Q^*Q}|\phi_i}<\ii$ in some orthonormal basis $(\phi_i)$ of the one-body space. Then $\tr(Q)=\sum_i\pscal{\phi_i|Q|\phi_i}$ is well-defined and does not depend on the chosen basis. In principle it is possible that the series $\sum_i\pscal{\phi_i|Q|\phi_i}$ converges for one specific basis even if the operator is not trace-class. This will be the case for our operator $P-\cP^0_-$.

Given an operator $Q$, we define $Q^{\epsilon\epsilon'}:=\cP^0_\epsilon Q\cP^0_{\epsilon'}$ where $\epsilon,\epsilon'\in\{\pm\}$ and $\cP^0_+:=1-\cP^0_-$. We say that an operator is $\cP^0_-$-trace class if $Q^{++}$ and $Q^{--}$ are trace-class and we define
$$\tr_0(Q):=\tr(Q^{++})+\tr(Q^{--})=\sum_i\pscal{\phi_i^+|Q|\phi_i^+}+\sum_i\pscal{\phi_i^-|Q|\phi_i^-}$$
for any chosen basis $(\phi_i^+)\cup(\phi_i^-)$ adapted to the decomposition induced by $\cP^0_-$. Of course if $Q$ is trace-class then it is also $\cP^0_-$-trace class but the converse is not true.

Now we remark that when $\tr(P-\cP^0_-)^2<\ii$ for a projector $P$, the operator $Q=P-\cP^0_-$ is automatically  $\cP^0_-$-trace class. The reason is that
$$(P-\cP^0_-)^2=(P-\cP^0_-)^{++}-(P-\cP^0_-)^{--}.$$
Additionally $\tr_0(P-\cP^0_-)$ is always an integer as proved in \cite[Lemma 2]{HLS1} and \cite{ASS}. We interpret
$e\tr_0(P-\cP^0_-)$ as the charge of the state $P$ (measured with respect to the free vacuum). Notice the condition $\tr(P-\cP^0_-)^2<\ii$ is a classical requirement of the Shale-Stinespring Theorem \cite{Thaller} which guarantees equivalence of Fock space representations.

Now we notice that when $P-\cP^0_-$ is $\cP^0_-$-trace class,
$$\tr_0(\cD^0(P-\cP^0_-))=\tr(|\cD^0|((P-\cP^0_-)^{++}-(P-\cP^0_-)^{--}))=\tr|\cD^0|(P-\cP^0_-)^2\geq0,$$
i.e. the kinetic energy is non negative and well defined when $\tr(P-\cP^0_-)^2<\ii$. Using Kato's inequality $|x|^{-1}\leq (\pi/2)|p|$ and \eqref{prop_D_0} we infer, following \cite{BBHS},
\begin{equation}
\frac{\alpha}{2}\iint\frac{|(P-\cP_-^0)(x,y)|^2}{|x-y|}dx\,dy\leq \frac{\pi\alpha}{4}\tr(|p|(P-\cP_-^0)^2)\leq \frac{\pi\alpha}{4}\tr_0(\cD^0(P-\cP^0_-)), 
\label{estim_exchange}
\end{equation}
i.e. the last term of \eqref{comput_BDF} is also well-defined.

Now we assume that 
$$V=-\alpha\nu\ast\frac{1}{|x|}$$
is the electrostatic potential created by a set of extended nuclei with (fastly decaying and smooth) total density $\nu$, $\int\nu=Z$. We define the BDF energy of $Q=P-\cP^0_-$ by 
$$\cE_{\rm BDF}^V(Q):=\tr_0(\cD^0Q)-\alpha D(\rho_Q,\nu)+\frac\alpha2 D(\rho_Q,\rho_Q)-\frac{\alpha}{2}\iint\frac{|Q(x,y)|^2}{|x-y|}dx\,dy$$
where
$$D(\rho,\rho'):=\iint\frac{\rho(x)\rho'(y)}{|x-y|}dx\,dy=4\pi \int\frac{\overline{\widehat{\rho}(k)}\widehat{\rho}'(k)}{|k|^2}dk$$
is the so-called Coulomb scalar product. 

It was proved in \cite[Lemma 1]{HLS3} that when $\tr Q^2<\ii$ and $Q$ is $\cP^0_-$-trace class, then $\rho_Q$ is a well-defined function which is squared-integrable and satisfies $D(\rho_Q,\rho_Q)<\ii$, hence $\cE_{\rm BDF}^V(Q)$ is well-defined by \eqref{estim_exchange}. Additionally we have when $0\leq\alpha\leq4/\pi$ by \eqref{estim_exchange} and using that $D(\cdot,\cdot)$ is a scalar product
$$\cE_{\rm BDF}^V(P-\cP^0_-)\geq -\frac\alpha2 D(\nu,\nu)>-\ii,$$
hence the BDF energy is bounded from below.

After these mathematical details, we are now able to minimize the BDF energy. We can either look for a \emph{global minimizer} which will be interpreted as the polarized vacuum in the presence of the external potential $V$, or for a \emph{minimizer with a charge constraint}
$$\tr_0(P-\cP^0_-)=N$$
which will usually represent the state of $N$ electrons coupled to the self-consistent polarized vacuum. We detail the two situations in the next sections. In both cases, the obtained minimizer will be $\cP^0_-$-trace class but not trace class (except when $V=0$), which will be related to renormalization as we will explain later.

\section{The Polarized Vacuum}
The polarized vacuum is by definition  the state of lowest QED energy in the Hartree-Fock no-photon class. By \eqref{comput_BDF}, it is also the state of lowest BDF energy. Hence we consider the following minimization problem
$$E^V:=\inf\cE_{\rm BDF}^V(P-\cP^0_-)$$
where the minimization is done over all orthogonal projectors $P$ acting on $\gH_\Lambda$ such that $P-\cP^0_-$ is $\cP^0_-$-trace class. As before the constraint on $P$ can be relaxed following Lieb \cite{Lieb} and replaced by the convex constraint $0\leq P\leq I$, but we do not detail this here.

It was proved in \cite{HLS1,HLS2} that a minimizer $P_{\rm vac}$ exists and that it solves the self-consistent equation
\begin{equation}
\left\{ \begin{array}{l}
P_{\rm vac}=\chi_{(-\ii;0]}\left(F_{P_{\rm vac}-I/2}\right),\\
F_{P_{\rm vac}-I/2}=D^0+\alpha(\rho_{[P_{\rm vac}-I/2]}-\nu)\ast\frac{1}{|x|}-\alpha\frac{(P_{\rm vac}-I/2)(x,y)}{|x-y|}.
\end{array}\right.
\label{SCF_pol_vac}
\end{equation}
Hence one more time the vacuum $P_{\rm vac}$ corresponds to filling negative energies of a self-consistent Fock operator. Notice 
\begin{equation}
 F_{P_{\rm vac}-I/2}=F_{0}+O(\alpha^2) \quad \text{where}\quad  F_{0}=D^0+V.
\label{def_Fock_pol_vac}
\end{equation}

\begin{figure}[h]
\input{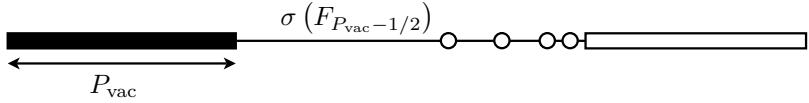}
\caption{The polarized vacuum $P_{\rm vac}$ in the presence of $V$ fills the negative energies of the SCF Fock operator $F_{P_{\rm vac}-I/2}$.}
\end{figure}

In general, one could have to create electron-positron pairs if one wants to deform $\cP^0_-$ into the polarized vacuum $P_{\rm vac}$. But when $V$ is not too strong it was proved in \cite{HLS2} that $P_{\rm vac}$ is unique and neutral:
$$\tr_0(P_{\rm vac}-\cP^0_-)=0.$$
In this case the vacuum $P_{\rm vac}$ only contains \emph{virtual} electron-positron pairs compared to $\cP^0_-$, see \cite[Appendix]{HLS3}.

In the right hand side of \eqref{def_Fock_pol_vac}, $\rho_{[P_{\rm vac}-I/2]}$, represents
the vacuum polarization density, which is self-consistently created
by the external potential $V$. 
Notice that one of the highlights our procedure is that although the
reference $\cP^0_-$ appears in the functional $\cE^V_{\rm BDF}$, the equation
\eqref{SCF_pol_vac} is independent of $\cP^0_-$, showing that the free vacuum
energy serves just as helpful device.

\section{Atoms and molecules}
For the study of common physical systems like atoms or molecules we have to consider the minimization of the BDF energy in charge sectors, that is to say imposing a constraint of the type 
$$\tr_0(P-\cP^0_-)=N$$
where $N\in\Z$.
Of course we cannot impose the number of particles but if $V$ is not too strong and $N>0$, this will provide a system of $N$ electrons interacting with the vacuum. Hence we introduce the following minimization problem
\begin{equation}
 E^V(N):=\inf_{\tr_0(P-\cP^0_-)=N}\cE_{\rm BDF}^V(P-\cP^0_-)
\label{def_E_V_N}
\end{equation}
where as before $P$ is assumed to be an orthogonal projector such that $P-\cP^0_-$ is $\cP^0_-$-trace class. It is not expected that a minimizer will always exist. If for instance $N$ is too large compared to the number $Z$ of nuclei, the system will certainly be unstable. On the other hand if $Z$ is too large, pairs could be created, which complicates the description of the system. In \cite{HLS3}, it was proved that when the following \emph{binding conditions} hold true
\begin{equation}
  E^V(N) < E^V(N-k)+E^0(k)\quad \forall k\in\Z\setminus\{0\},
\label{binding}
\end{equation}
then a minimizer exists for $E^V(N)$. The binding condition \eqref{binding} was proved to hold true in \cite{HLS3} when for instance $0\leq N\leq Z$ and $\alpha\ll1$ (non relativistic limit).

Here we assume that there is a minimizer $P$ and that $V$ is not too strong. Then it was proved in \cite{HLS3} that $P$ solves the SCF equation
\begin{equation}
\left\{ \begin{array}{l}
P=\chi_{(-\ii;\mu]}\left(F_{P-I/2}\right),\\
F_{P-I/2}=D^0+\alpha(\rho_{[P-I/2]}-\nu)\ast\frac{1}{|x|}-\alpha\frac{(P-I/2)(x,y)}{|x-y|}
\end{array}\right.
\label{SCF_mol}
\end{equation}
where $\mu$ is a Fermi level (a Lagrange multiplier due to the charge constraint). We can write
$$P=P_{\rm vac}+P_{\rm el}$$
where
$$P_{\rm vac}=\chi_{(-\ii;0]}\left(F_{P-I/2}\right)\quad \text{and}\quad P_{\rm el}=\chi_{(0;\mu]}\left(F_{P-I/2}\right)=\sum_{i=1}^N|\phi_i\rangle\langle\phi_i|$$
with 
$$F_{P-I/2}\phi_i=\lambda_i\phi_i$$
for all eigenvalues $\lambda_i\leq\mu$. The orbitals $(\phi_i)_{i=1}^N$ describe the Hartree-Fock state of the $N$ electrons whereas $P_{\rm vac}$ describes the SCF polarized vacuum in the presence of the external field $V$ and the $N$ electrons.

\begin{figure}[h]
\input{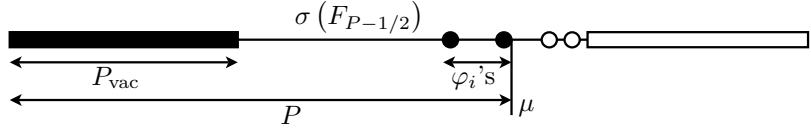}
\caption{Decomposition of the system `vacuum + $N$ electrons' for the solution $P$ in the $N$th charge sector.}
\end{figure}

We notice that the decomposition of the state $P$ into $N$ electrons and the polarized vacuum can be made unambiguous because $P$ satisfies the SCF equation \eqref{SCF_mol}. For a general state $P$ satisfying $\tr_0(P-\cP^0_-)=N$, there is no canonical decomposition between real and virtual particles.

Now we remark that
\begin{equation}
 F_{P-I/2}=F_{\rm el}+\alpha\rho_{[P_{\rm vac}-I/2]}\ast\frac{1}{|x|}-\alpha\frac{(P_{\rm vac}-I/2)(x,y)}{|x-y|}=F_{\rm el}+O(\alpha^2)
\label{decomp_F_mol}
\end{equation}
where 
$$F_{\rm el}=D^0+\alpha(\rho_{P_{\rm el}}-\nu)\ast\frac{1}{|x|}-\alpha\frac{P_{\rm el}(x,y)}{|x-y|}$$
is the usual Dirac-Fock operator for $N$ relativistic electrons. Hence we deduce that the $\phi_i$'s solve the usual Dirac-Fock equations \cite{Sw,ES}, perturbed by the SCF vacuum polarization potentials. An essential feature is of course that these equations have been obtained by a \emph{minimization procedure}, contrarily to the usual Dirac-Fock case. 

Notice the Dirac-Fock model is \emph{not} obtained as a \emph{variational} approximation of the BDF model. But the Dirac-Fock equations are an approximation of the BDF equations. This was first noted by Chaix and Iracane in \cite{CI}. 

\section{Time-dependent equation}
The time-dependent equation corresponding to our model could also be useful, in particular for the study of spontaneous pair creation which is usually formulated as an adiabatic theory on the evolution equation \cite{Nenciu,RGA}. It reads
$$i\dot{P}(t)=[F_{[P(t)-I/2]},P(t)]$$
where we choose as initial condition an orthogonal projector $P(0)$ such that $\tr(P(0)-\cP^0_-)^2<\ii$. It was proved in \cite{HLSp} that this equation admits a global-in-time solution $P(t)$, $t\in\R$, which has a constant BDF energy and charge:
$$\forall t\in\R,\quad \cE^V_{\rm BDF}(P(t)-\cP^0_-)=\cE^V_{\rm BDF}(P(0)-\cP^0_-),\quad \tr_0(P(t)-\cP^0_-)=\tr_0(P(0)-\cP^0_-).$$

\section{Renormalization}
In regular QED, the divergences of the (appropriately defined) physical measurable quantities are usually eliminated by means of a \emph{mass} and a \emph{charge renormalization}. The main idea is to assume that the parameters $\alpha$ and $m$ appearing in the theory are indeed \emph{bare parameters} which are not physically observable. The \emph{physical parameters} are assumed to be functions of $\alpha$, $m$ and the cut-off $\Lambda$
$$\alpha_{\rm ph}=\alpha_{\rm ph}(\alpha,m,\Lambda),\qquad  m_{\rm ph}=m_{\rm ph}(\alpha,m,\Lambda)$$
and equal the physical values obtained in experiment. These functions should be inverted in order to express the unknown bare quantities in term of the physical quantities
\begin{equation}
\alpha=\alpha(\alpha_{\rm ph},m_{\rm ph},\Lambda),\qquad  m=m(\alpha_{\rm ph},m_{\rm ph},\Lambda).
\label{def_bares_quantities}
\end{equation} 
Using these functions, one expects to
remove (in some sense that needs to be precised) all divergences from physically measurable quantities.

 Mass and charge renormalization however does not
remove all divergences in the theory. Certain quantities, e.g. the
bare Feynman propagator $S_F$ (either at equal times or at general space
time points), are still divergent. The expectation is that all these
divergences cancel in physically measurable quantities and that they are
therefore of no real relevance in formulating the theory.

Although there is no real need to do this, it is often convenient to
introduce a renormalization of the bare Feynman propagator $S_F$. This is
referred to as a \emph{wavefunction renormalization}. In full QED \cite{Dy2} it is claimed
that the divergence in the Feynman propagator may be removed by a
\emph{multiplicative renormalization} and that the renormalized propagator has
the same pole near mass shell in 4-momentum space as a free propagator
corresponding to a particle with the correct physical mass.

Note that in practice, this theoretical renormalization procedure is always used to justifying the dropping of the divergent terms obtained at each order of the perturbation theory \cite{Dy2}. For this fact to be true, it is particularly important that renormalization can be expressed by means of \emph{multiplicative} parameters in front of the different propagators \cite{Dy2}.

In Hartree-Fock QED, it is not clear at all if the usual renormalization program of QED can be applied, especially when photons are not included. In  \cite[p. 194--195]{RGA}, it is argued that mass and charge renormalization is alone not enough to completely remove the divergences of the HF theory by means of multiplicative parameters.

In any case, the physical mass and charge have to be identified within the model. We propose the following definitions. 
The physical mass is just the lowest energy of a free electron, hence
\begin{equation}
 m_{\rm ph}(\alpha,m,\Lambda):=E^0(1)
\label{m_ph}
\end{equation}
which was defined in \eqref{def_E_V_N}.

To define the physical coupling constant, we consider an extended nucleus of density $\nu$, $\int\nu=Z$, and put it in the vacuum. Let $Q_{\rm vac}=P_{\rm vac}-\cP^0_-$ be the polarized vacuum solution of \eqref{SCF_pol_vac}. We assume that $\nu$ is not too strong such that the vacuum stays neutral, $\tr_0(Q_{\rm vac})=0$. Of course in reality it is impossible to distinguish the nucleus from the vacuum and the charge which is observed far way from the nucleus is just
$$e\left(Z-\int_{\R^3}\rho_{Q_{\rm vac}}\right)$$
(provided $\rho_{Q_{\rm vac}}$ is an $L^1$ function). Hence we may define 
\begin{equation}
 \alpha_{\rm phys}(\alpha,m,\Lambda):=\alpha\left(1-Z^{-1}\int_{\R^3}\rho_{Q_{\rm vac}}\right).
\label{alpha_ph}
\end{equation}
If the above formula still depends on $Z$, one can take the limit as $Z\to0$.

It is very important to realize that charge renormalization is based on the fact that the operator $Q_{\rm vac}$ is \emph{not} trace-class. If it were trace-class, one would of course have
$\tr_0(Q_{\rm vac})=0=\int\rho_{Q_{\rm vac}}$, hence $\alpha_{\rm ph}=\alpha$. Therefore, the mathematical difficulty that a minimizer of the BDF energy is never trace-class (except when $\nu=0$) is the origin of charge renormalization. Also this shows that in a finite dimensional space (for computational purpose for instance), renormalization is certainly more involved as all operators are trace class.

Both \eqref{m_ph} and \eqref{alpha_ph} would define $m_{\rm ph}$ and $\alpha_{\rm ph}$ as extremely complicated non-linear functions of $\alpha$, $m$ and $\Lambda$. A challenging task is to study the finiteness of measurable quantities like for instance the energy of an electron in the presence of an external field $E^V(1)$, when $\alpha_{\rm ph}$ and $m_{\rm ph}$ are fixed to be the observed physical quantities. We do not know if this is possible when photons are not taken into account.

\bigskip

It is however possible to completely solve the above program for a (further) simplified model called the \emph{reduced} Hartree-Fock, as was done in \cite{HLS2}. We explain that now.

The reduced HF model is just obtained by neglecting the exchange term in the HF energy \eqref{HF_QED}
\begin{equation}
 \cE_{\rm rHF}^V(\gamma)=\tr(D^0\gamma)+\int V(x)\rho_\gamma(x)\,dx
+\frac{\alpha}{2}\iint\frac{\rho_\gamma(x)\rho_\gamma(y)}{|x-y|}dx\,dy.
\label{rHF_QED}
\end{equation}
This is natural as the exchange term is usually treated together 
with a term describing the interaction with
the photon field to form the standard electron
{\em self-energy} that is a subject of the mass renormalization. 

The so-obtained  model is much simpler than the HF model as the energy is now a convex function of $\gamma$. All what we have said concerning the case with exchange term can be extended to this simplified model. The free vacuum is even a simpler object as in Equation \eqref{SCF_free_vac} only the exchange term created a self-consistent field. Hence we obtain
$$\cP^0_-=P^0_-\quad \text{and}\quad \cD^0=D^0.$$
The \emph{reduced Bogoliubov-Dirac-Fock} (rBDF) energy then reads \cite{HLS2}
\begin{equation}
 \cE_{\rm rBDF}^V(Q)=\tr_0(D^0Q)-\alpha D(\rho_Q,\nu)+\frac\alpha2 D(\rho_Q,\rho_Q).
\label{def_rBDF}
\end{equation}
It can easily be shown that for a free electron in the vacuum \cite[Lemma 3]{HLS3}
$$\inf_{\tr_0(P-P^0_-)=1}\cE^0_{\rm rBDF}(P-P^0_-)=m,$$
 i.e. $m_{\rm ph}=m$ and there is no mass renormalization for the reduced BDF model.

Consider now a small external density $\nu$, $\int\nu=Z$ and let $Q_{\rm vac}$ be the associated polarized vacuum, with density $\rho_{\rm vac}:=\rho_{Q_{\rm vac}}$. The SCF equation satisfied by $Q_{\rm vac}$ reads
\begin{equation}
 Q_{\rm vac}=\chi_{(-\ii;0]}(F)-P^0_-
\label{SCF_rBDF}
\end{equation}
where 
$$F=D^0 + \alpha(\rho_{\rm vac}-\nu) \ast \frac 1{|x|}.$$
We expand \eqref{SCF_rBDF} in powers
of $\alpha$, using that $0\notin\sigma(F)$ when $\nu$ is small enough. We can use the resolvent representation \cite[Section VI, Lemma 5.6]{K} to
derive the self-consistent equation for the density $\rho_{\rm vac}$ 
\begin{equation}
\rho_{\rm vac}(x) = -\frac 1{2\pi} \dK \, \Tr \left[ \frac 1{D^0 + \alpha(\rho_Q-\nu) \ast \frac 1{|x|}
+ i\eta}
- \ide \right](x,x).
\end{equation}
Applying the resolvent equation $$ \frac 1{A-\alpha B} - \frac 1{A} = \alpha \frac 1{A} B \frac 1{A}
+ \alpha^2 \frac 1{A} B \frac 1{A} B \frac 1{A} + \alpha^3 \frac 1{A} B \frac 1{A} B \frac 1{A}
B \frac 1{A-\alpha B}$$
and using
Furry's Theorem \cite{Furry}, telling us that the corresponding $\alpha^2$-term with two potentials
vanish, we obtain
\begin{equation}\label{rscer}
\rho_{\rm vac} = \alpha F_1[\rho_{\rm vac} - \nu] +   F_{3}[\alpha(\rho_{\rm vac} - \nu)],
\end{equation}
\begin{multline*}
F_{3} [\rho](x) =\frac1{2\pi}\times \\ \dK\, \Tr \left[\ide \rho\ast \frac 1{|x|} \ide \rho\ast \frac 1{|x|}
\ide\rho\ast \frac 1{|x|} \frac 1{D^0 + \rho\ast\frac{1}{|x|} + i\eta }\right](x,x).
\end{multline*} 

As realized first by Dirac \cite{Dir,D2} and Heisenberg \cite{Hei}, cf. also \cite{FO},
the term $F_1[\rho]$ plays a particular role since it is
logarithmically ultraviolet divergent. Following, e.g., Pauli-Rose \cite{PauliRose},
one evaluates in Fourier representation $$\widehat F_1[\rho](k)=-\hat\rho(k)B_\Lambda(k),$$ 
where \cite[Eq. (5)--(9)]{PauliRose}
$ B_\Lambda(k)= B_\Lambda -  C_\Lambda(k)$, with
\begin{equation}
B_\Lambda = B_\Lambda(0) = \frac 1\pi \int_0^{\frac{\Lambda}{\sqrt{1+\Lambda^2}}}\frac{z^2-z^4/3}{1-z^2}\,dz
=\frac{2}{3\pi}\log(\Lambda)-\frac{5}{9\pi}+\frac{2}{3\pi}\log 2 + O(1/\Lambda^2) .
\end{equation}
and 
\begin{equation}
 \lim_{\Lambda \to \infty} C_{\Lambda}(k) = C(k)=-\frac 1{2\pi} \int_0^1 dx (1-x^2)\log[1+k^2(1-x^2)/4],
\end{equation}
which was first calculated by Serber and Uehling \cite{Se,Ue}. 

We can now compute the physical coupling constant. First we rewrite \eqref{rscer} in Fourier space as
\begin{equation}\label{rscer2}
(1+\alpha B_\Lambda)\widehat{\rho}_{\rm vac}(k) = \alpha B_\Lambda\widehat{\nu}(k)+\alpha C_\Lambda(k)(\widehat\rho_{\rm vac} - \widehat\nu)(k) +   \widehat{F}_{3}[\alpha(\rho_{\rm vac} - \nu)](k).
\end{equation}
Assuming that $\rho_{\rm vac}\in L^1(\R^3)$ and taking $k=0$, we find
$$\int\rho_{\rm vac}=\frac{\alpha B_\Lambda Z}{1+\alpha B_\Lambda}\neq0$$
where we have used that $C_\Lambda(0)=\widehat{F}_3[\alpha(\rho_{\rm vac} - \nu)](0)=0$. Hence by \eqref{alpha_ph} we find
\begin{equation}
\boxed{\alr = \frac \alpha{1+\alpha B_\Lambda}.}
\label{ren_alpha}
\end{equation} 
It follows that
necessarily $\alr B_\Lambda < 1$. We emphasize that although in the literature the expression of $\alr$ is sometimes expanded to get $\alr\simeq \alpha(1-\alpha B_\Lambda)$ leading to the condition $\alpha B_\Lambda<1$, the real constraint indeed applies to the physically observed $\alr$ and not the bare one.

We now show how to renormalize the SCF equation using \eqref{ren_alpha}. Denote $\rho= \rho_{\rm vac} - \nu$ the total (observable) density, then
\eqref{rscer} can be rewritten in terms of $\rho$
\begin{equation}\label{r1}
\alpha\widehat \rho =-\alpha\widehat\nu  -\alpha^2 B_\Lambda \widehat \rho  + \alpha^2
C_\Lambda (k) \hr +  \alpha\widehat F_{3} [\alpha\rho]
\end{equation}
and
\begin{equation}\label{r2}
\alpha\hat \rho =-\frac \alpha{1+\alpha B_\Lambda } \widehat\nu   + \frac \alpha{1+\alpha B_\Lambda}
C_\Lambda (k)\alpha \hr + \frac \alpha{1+\alpha B_\Lambda } \widehat F_{3} [\alpha\rho].
\end{equation}
To perform our renormalization scheme we fix as physical
(renormalized) objects $\alr\rr = \alpha \rho$. Notice the renormalization of the density $\rho$ is similar to a wavefunction renormalization of the (equal time) Feynman propagator as explained above. We can
rewrite the self-consistent
equation \eqref{r1} as
\begin{equation}\label{r3}
\alr\widehat \rho_{\mathrm{ph}} =-\alr \widehat\nu   +  \alpha_{\mathrm{ph}}^2
C_\Lambda (k) \widehat \rho_{\mathrm{ph}} +  \alpha_{\mathrm{ph}} \widehat F_{3} [\alr \rr],
\end{equation}
independently of  the bare $\alpha$.  
Notice that equation \eqref{r3} satisfied by $\alpha_{\rm ph}\rho_{\rm ph}$ is exactly the same as equation \eqref{r1} satisfied by $\alpha\rho$, but with the logarithmically divergent term $\alpha^2 B_\Lambda \widehat \rho$ dropped. Therefore, as usual in QED \cite{Dy2}, the charge renormalization allows to simply justify the dropping of the divergent terms in the self-consistent equation. In practice \cite{MPS}, one would solve \eqref{r3} with $\alpha_{\rm ph}\simeq 1/137$ and with $C_\Lambda(k)$ replaced by its limit $C(k)$.

Returning to the {\em effective} Hamiltonian $F=D^0  +\alpha (\rho_{\rm vac}-\nu)\ast1/|x|$ and
inserting \eqref{r3}, i.e.
expressing in terms of the physical
objects, we obtain
\begin{equation}\label{effd}
D^0 + \alr \rr\ast \frac 1{|x|} = D^0 - \alr \nu\ast\frac 1{|x|}
+V_{{\rm eff}},
\end{equation}
with 
$$V_{{\rm eff}} =\frac 2{\pi^3} \mathcal{F}^{-1}\left[\frac {\alpha_{\mathrm{ph}}^2 C_\Lambda (k)\widehat 
\rho_{\mathrm{ph}} (k)
+ \alr \hat F_3(\alr \rr)}{ k^2}\right](x)$$
the effective self-consistent potential, where $\mathcal{F}^{-1}$ denotes the inverse Fourier transform.
Notice, this equation is valid for any strength of the external potential.
However, expanding $\rr$ in $\alr$, 
we obtain to lowest order in $\alr$
\begin{eqnarray*}
 V_{{\rm eff}} &\simeq&
\alpha_{\mathrm{ph}}^2 \frac 2{\pi^3} \cF^{-1}\left[\frac {C_\Lambda (k)\widehat\nu (k)}{ k^2}\right](x) \\
&\simeq&
\frac {\alpha_{\mathrm{ph}}^2 }{3\pi} \int_1^\infty dt (t^2-1)^{1/2}\left[\frac 2{t^2} + \frac 1{t^4}\right] \int dx'
e^{-2|x-x'|t} \frac {\nu(x')}{|x-x'|},
\end{eqnarray*}
the Uehling potential \cite{BR}. 

\section{The Landau Pole}
We notice that \eqref{ren_alpha} can be written as
$$\alpha=\frac{\alpha_{\mathrm{ph}}}{1-\alpha_{\mathrm{ph}} B_\Lambda}.$$
The fact that the denominator can go to zero is usually called the Landau pole. Also we see that 
\begin{equation}
 \alr B_\Lambda < 1
\label{cond_alpha_phys}
\end{equation}
 which proves that $\alr\to0$ when $\Lambda\to\ii$, independently of $\alpha$. 

In \cite[Thm 2]{HLS2}, it was proved that for a fixed (and not too strong) external field $V=-\alpha\nu\ast\frac1{|x|}$, the unique polarized vacuum $P_\Lambda$ of the \emph{reduced} BDF model satisfies
$$\lim_{\Lambda\to\ii}\tr\left(P_\Lambda-P^0_-\right)^2=0\quad \text{and}\quad \lim_{\Lambda\to\ii}D(\rho_{P_\Lambda-P^0_-}-\nu,\rho_{P_\Lambda-P^0_-}-\nu)=0.$$

In words, when $\Lambda\to\infty$, the vacuum polarization density \emph{totally cancels the external density $\nu$}, for $\rho_{P_\Lambda-P^0_-}\to\nu$. But since $P_\Lambda-P^0_-\to0$, this means that in the limit $\Lambda\to\infty$, $P_\Lambda-P^0_-$ and its associated density become independent. Therefore, the minimization without cut-off makes no sense both from a mathematical and physical point of view. Indeed all this easily implies that when no cut-off is imposed and when $\nu\neq0$, the infimum of the reduced BDF functional is \emph{not attained}.
In physics, this ``nullification" of the theory as the cut-off $\Lambda$ diverges has been first suggested by Landau {\it et al.} \cite{Lan86,Lan84,Lan89,Lan100} and later studied by Pomeranchuk {\it et al.} \cite{Pom}.

We notice that with the usual value $\alpha_{\rm ph}\simeq \frac{1}{137}$, \eqref{cond_alpha_phys} leads to the physical bound $\Lambda< 10^{280}$ (in units of $mc^2$).

\section{Conclusion}
We have presented a model which is obtained as the mean-field approximation of no-photon QED. We believe that the Hartree-Fock approximation is an interesting model as it possesses already many peculiarities of the full QED and it is much simpler to handle. In particular, optimized states always correspond to filling the spectrum of a one-body operator up to some Fermi level $\mu$, which corresponds to the original interpretation of Dirac.

The main advantage of this model is that it is \emph{variational}: states can be found by minimizing an energy, contrarily to the usual relativistic effective models used for instance in Quantum Chemistry. This provides a better interpretation of the optimal states. Also the model provides a justification of the Dirac-Fock equations, which are seen as a $O(\alpha^2)$ approximation of a set of equations obtained by minimization.

Another advantage of the model is that it is \emph{nonperturbative}: the only constraint to have a globally stable model is that $0\leq\alpha< 4/\pi$. The equations are quite simple and renormalization can be done non perturbatively to all orders (at least when the exchange term is neglected).

The main idea in the derivation of our model was first to define the SCF free vacuum by a thermodynamic limit, and then to subtract its infinite energy in order to get a bounded-below function. This method replaces the usual normal-ordering which can only be used for non interacting systems. In principle the same method could be used for the full QED. But probably it is not possible to express the difference between the energy of the considered state and the one of the free vacuum in a simple way. 

We have neglected photons but in principle one could take the photon field into account. The mathematical study of such a theory remains to be done.

\bigskip

\noindent\textbf{Acknowledgment.} M.L. and E.S. acknowledge support from the ANR project ``ACCQUAREL'' of the French ministry of research.


\bibliographystyle{amsplain}

\end{document}